\documentclass{article}
\usepackage{spconf,amsmath,graphicx}

\usepackage{xcolor}  

\usepackage{booktabs}
\usepackage{multirow}
\usepackage{float}
\usepackage{makecell}
\usepackage{pifont}
\usepackage{hyperref}
\usepackage{subcaption}
\usepackage[export]{adjustbox}
\newcommand{\cmark}{\ding{51}}%
\newcommand{\xmark}{\ding{55}}%

\newcommand{\ian}[1]{{\color{purple}[Ian: #1]}}
\newcommand{\hc}[1]{{\color{blue}[HC: #1]}}

\title{Integrating self-supervised speech model with pseudo word-level targets from visually-grounded speech model}
\name{
\parbox{\linewidth}{
\centering
Hung-Chieh Fang$^{1*}$, Nai-Xuan Ye$^{1*}$\thanks{$^*$ Equal contribution.}, Yi-Jen Shih$^{1,2}$, Puyuan Peng$^2$, Hsuan-Fu Wang$^1$,  Layne Berry$^2$, Hung-yi Lee$^1$, David Harwath$^2$ 
}
}
\address{
  $^1$ National Taiwan University, Taiwan \\
  $^2$ The University of Texas at Austin, USA}

\begin{document}
\ninept
\maketitle
\begin{abstract}
    Recent advances in self-supervised speech models have shown significant improvement in many downstream tasks. However, these models predominantly centered on frame-level training objectives, which can fall short in spoken language understanding tasks that require semantic comprehension. Existing works often rely on additional speech-text data as intermediate targets, which is costly in the real-world setting. 
    To address this challenge, we propose Pseudo-Word HuBERT (PW-HuBERT), a framework that integrates pseudo word-level targets into the training process, where the targets are derived from a visually-ground speech model, notably eliminating the need for speech-text paired data. 
    Our experimental results on four spoken language understanding (SLU) benchmarks suggest the superiority of our model in capturing semantic information.
\end{abstract}


\vspace{-5pt}
\section{Introduction}
\vspace{-5pt}
\label{sec:intro}
Traditional supervised learning methods demand a substantial amount of labeled data, which is challenging to acquire in real-world scenarios, especially in speech processing. 
Recent research \cite{mohamed2022self} has shifted the focus to self-supervised learning (SSL), which pre-trained an upstream model with unlabeled data to obtain the representations. 
These representations are subsequently leveraged to train downstream models for tasks with only a limited amount of labeled data. 

One branch of SSL study \cite{chrupala2022visually} on speech is visually-grounded speech (VGS) models, where the models are trained with speech-image paired data to boost semantic information. 
VGS models have shown success in a wide range of tasks such as speech recognition \cite{hsu2019transfer,shi2022learning}, speech generation \cite{hsu-etal-2021-text}, syllable and word discovery \cite{harwath2019towards,peng2023syllable, peng2022word}, keyword localization \cite{Leanne_keyword}, and speech-image retrieval \cite{Sanabria2021TalkDW,peng2022fast,shih2023speechclip, berry2023m, bhati23_interspeech}. 
The recently proposed VG-HuBERT \cite{peng2022word} is a VGS model that reaches state-of-the-art on the word segmentation task. 
For a given unlabeled utterance, a simple threshold is applied to the Transformer layer's attention weights to predict word segments. 
A K-Means clustering algorithm is further applied for segment categorization.
Although VG-HuBERT has been shown to contain rich word boundary information in its representation \cite{peng2023syllable}, our experimental results show its limited performance on general SLU tasks.

While recent state-of-the-art SSL speech models \cite{hsu2021hubert, baevski2020wav2vec} achieve remarkable performance on various downstream tasks, it is worth noting that their training objectives are primarily frame-level. 
However, this may pose a challenge for SLU tasks, as such tasks often require more coarse-grained, word-level information. 
To alleviate the issue, previous works introduce a framework comprised of an acoustic speech recognition (ASR) model and a natural language understanding (NLU) model. However, such an approach often requires paired transcripts, which are expensive to obtain. 
A recent work \cite{wu2023improving} proposes a textless approach that utilizes discrete units as intermediate targets to overcome the limitation. Notably, this method is specifically tailored for downstream models. 

In this paper, we investigate the question of capturing semantic information in self-supervised speech models. 
We propose Pseudo-Word HuBERT (PW-HuBERT), a self-supervised model that is trained with pseudo word-level targets.
In particular, we leverage word boundaries from word-segmentation models to generate pseudo word-level targets and apply them to HuBERT \cite{hsu2021hubert} pretraining. Our motivation is that word boundaries can aid foundation models in understanding the relationships between tokens within the same word. In contrast to the original HuBERT, where each target is independently assigned based on a fixed frequency, thus lacking contextual association.



We introduce two different architectures: single and hierarchical. 
The architecture of our single PW-HuBERT is similar to HuBERT.
We add two extra transformer layers to predict the pseudo word-level targets generated from VG-HuBERT \cite{peng2022word}. 
For the hierarchical PW-HuBERT, we apply the original HuBERT pretraining objective as an additional frame-level loss to further guide the training process. 
We evaluate our models on multiple SLU benchmarks and downstream tasks, including SLUE, SLUE Phase-2, SNIPS, and ZeroSpeech 2021 semantics, showing great improvement in the experimental results.
\begin{figure*}[ht]
\centering
    \begin{subfigure}[t]{0.33\textwidth}
    \centering
    \includegraphics[width=0.83\textwidth, trim={0cm 0.1cm 0cm 0cm}]{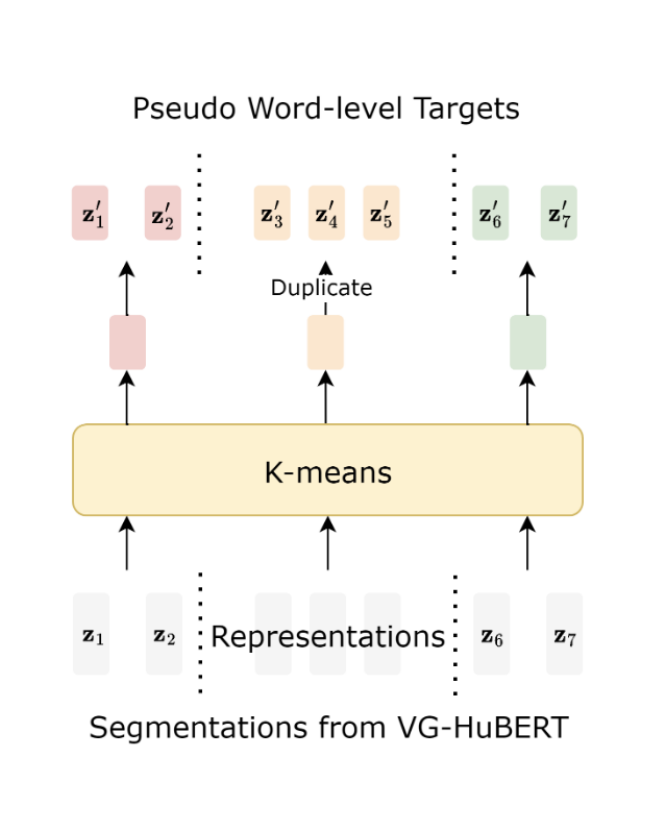}
    \caption{Pseudo word-level target generation} \label{fig:gen}
\end{subfigure}\hfill
\begin{subfigure}[t]{0.33\textwidth}
    \centering
\includegraphics[width=\textwidth, trim={0cm 0cm 0cm 1cm}]{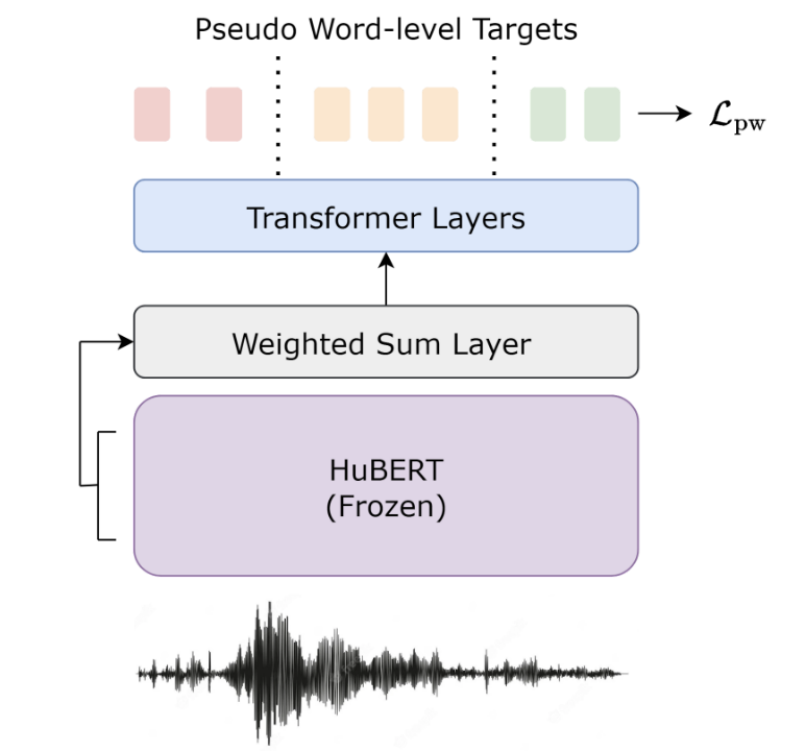}
    \caption{Single PW-HuBERT} \label{fig:single}
\end{subfigure}\hfill
 \begin{subfigure}[t]{0.33\textwidth}
    \centering
    \includegraphics[width=\textwidth, trim={0cm 0cm 0cm 1cm}]{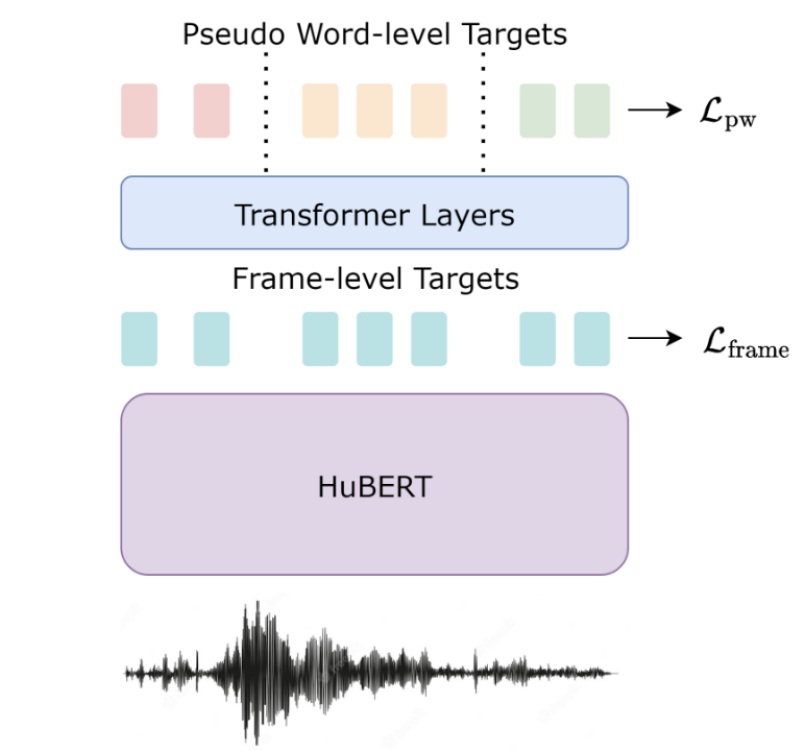}
    \caption{Hierarchical PW-HuBERT} \label{fig:hierarchical}
\end{subfigure}

\caption{Overview of pseudo word-level target generation and PW-HuBERT models. (a) With the word segmentations from VG-HuBERT~\cite{peng2022word}, we apply mean-pooling on the representations within the same segment. A K-means model is used to predict the cluster ID of each segment, which is then duplicated to match the length of the corresponding segment. (b) The HuBERT weights are frozen, and the features from each layer are passed to a learnable weighted sum layer to get the final representation. The representation is then passed to another two transformer layers to predict pseudo word-level targets. (c) Hierarchical PW-HuBERT predicts frame-level targets at the 12\textsuperscript{th} layer and predicts pseudo word-level targets at the 14\textsuperscript{th} layer.}
\label{fig:overview}
\end{figure*}


Our contributions can be summarized as follows:
\begin{itemize}
    \setlength\itemsep{0.1em}
    \item We are the first work to leverage an unsupervised word discovery model for speech SSL model pre-training.
    \item We demonstrate the benefits of joint training with frame-level and word-level units for capturing semantic information.
    
    
\end{itemize}



\vspace{-15pt}
\section{Method}
\vspace{-5pt}
\label{sec:method}
\subsection{Preliminaries}
\textbf{Hidden-unit BERT (HuBERT) \cite{hsu2021hubert}.}  HuBERT is a self-supervised model that simulates masked language modeling, where their targets are generated by clustering on acoustic features to provide frame-level supervision. 
HuBERT produces good feature representations for speech downstream tasks. 
In this paper, we use HuBERT-Base as our backbone model, which includes a CNN feature extractor and 12 transformer layers. 

\noindent\textbf{Visually-grounded HuBERT (VG-HuBERT) \cite{peng2022word}.} VG-HuBERT is trained with a generic image-speech contrastive loss, which exhibits state-of-the-art word discovery capabilities by applying a binary threshold to the self-attention map.

\vspace{-15pt}
\subsection{Word-Boundary and Pseudo Target Generation}
\label{sec:target}
In this section, we demonstrate our utilization of unsupervised word segmentation to generate pseudo word-level targets. Specifically, we leverage the state-of-the-art method VG-HuBERT for word discovery. It is worth highlighting that this approach can potentially be extended to various unsupervised word segmentation methods.

Fig.~\ref{fig:gen} illustrates the process of generating pseudo word-level targets. Given a speech utterance $\mathbf{X} = [\mathbf{x}_1, \dots, \mathbf{x}_T] $ of $T$ frames, we first apply VG-HuBERT to generate the feature sequence $\mathbf{Z} = [\mathbf{z}_1, \dots, \mathbf{z}_T]$ and the attention boundaries $\mathbf{B} = [(s_1, e_1), \dots, (s_u, e_u)]$, where $u$ is the number of segments. 
Following the previous work \cite{peng2022word}, we do not directly use the attention segments as they tend to be narrower than the actual word segments.
Instead, we use the midpoints between the adjacent attention boundaries as our pseudo word boundaries $\mathbf{B'} = [(s_1', e_1'), \dots, (s_u', e_u')]$, where $s_i' = \frac{e_{i-1} + s_i}{2}$ and $e_i' = \frac{e_{i} + s_{i+1}}{2}$. (The first and the last stay the same: $s_1'=s_1,e_u'=e_u$)

We then apply mean pooling on features within the same segment to construct the pseudo word features \footnote{Note that it does not necessarily map to real words; instead, it represents higher-level features compared to the original HuBERT.}$\mathbf{W} = [\mathbf{w}_1, \dots ,\mathbf{w}_u]$, where $\mathbf{w}_i = \mathrm{mean}(\mathbf{z}_{s_i}, \dots, \mathbf{z}_{e_i})$. 
After that, We run the K-means clustering algorithm on $\mathbf{W}$ and assign cluster ID 
 $\mathbf{z_i'}$ to each $\mathbf{w}_i$.
In order to match the sequence length $T$ of the input speech utterance, we simply duplicate each cluster ID to its corresponding length. e.g. the cluster ID for $\mathbf{w}_i$ will be duplicated $e_i-s_i+1$ times.
Lastly, we concatenate all the cluster IDs and form the pseudo word-level target sequence $\mathbf{Z'} = [\mathbf{z_1'}, \dots, \mathbf{z_T'}]$
\subsection{Single PW-HuBERT}
\label{method:single}
To distill word-level information in the HuBERT model, we propose Single PW-HuBERT illustrated in Fig.~\ref{fig:single}.
Specifically, we weighted-sum the representations in each layer of HuBERT (same as \cite{shih2023speechclip, yang21c_interspeech}) and feed it into two additional learnable transformer layers to learn the word level semantics.
The HuBERT model is frozen to reduce computational cost and increase the training stability. 



Following the training procedure of HuBERT, we define $\hat {\mathbf{X}}$ as the masked version of input speech sequence 
$\mathbf{X}$ and $\mathbf{M}$ as the masked indices generated randomly, where embeddings of $\hat {\mathbf{X}}$ at timestep $t \in \mathbf{M}$ are replaced with mask embedding. 
Our model $f$ has to reconstruct $\mathbf{X}$ from $\hat{\mathbf{X}}$ by predicting probability $f(\cdot | \hat{ \mathbf{X}}, t)$ over each masked timestep. 
We then calculate cross-entropy loss between target tokens $\mathbf{Z'}$ 
generated in section~\ref{sec:target} and the model's prediction as $\mathcal{L}_{\mathrm{pw}}$, which is defined as:
\begin{equation}
\mathcal{L}_{\mathrm{pw}} = \sum_{t \in \mathbf{M}} \log f(\mathbf{z_t'} | \hat {\mathbf{X}}, t)    
\end{equation}
Following the original HuBERT \cite{hsu2021hubert}, we only calculate the loss on masked segments to be more resilient.

Overall, to retain the performance of HuBERT while incorporating word-level targets into the training objective to capture semantic information, we follow the pre-training convention~\cite{shih2023speechclip, yang21c_interspeech} to freeze HuBERT layers, and pre-trained two more layers on top of it with pseudo-word labels.

\begin{table*}[ht]
\centering
\caption{Performance of different models on SLUE, SLUE-2, ZeroSpeech 2021 semantics track, and SNIPS. $^\dagger$ indicates that the model is trained with a similar task setup to ZeroSpeech semantics.}
\resizebox{0.9\linewidth}{!}{
\begin{tabular}{c|cc|c|cc|cc}
\toprule
\multirow{2}{*}{\bf Dataset} & \multicolumn{2}{c|}{\bf SLUE} & \bf SLUE Phase-2 & \multicolumn{2}{c|}{\bf SNIPS} & \multicolumn{2}{c}{\bf ZeroSpeech 2021 semantics}
\\
 & \bf SA & \bf NER & \bf NEL & \bf SF  & \bf IC & \bf librispeech & \bf synthetic \\
\midrule
 Metric &  F1 $\uparrow$ & F1 / Label F1 $\uparrow$ & Frame F1 / Word F1 $\uparrow$  &  \multicolumn{2}{c|}{F1 $\uparrow$} & \multicolumn{2}{c}{ Similarity Judgement $\uparrow$ } \\
\midrule 

HuBERT \cite{hsu2021hubert} & 45.27 & 51.6 / 64.8 & 57.54 / 61.14  & 88.16  & 98.57 & 5.71 & 6.79  \\
HuBERT$_{14}$ & 44.54 & 51 / 66.8 & 58.43 / 61.84 & 88.18 & \underline{98.71} & 5.11 & 6.63   \\
VG-HuBERT \cite{peng2022word} & 45.1 & 41.4 / 52.3 & 47.11 / 51.52 & 84.98 & 98.42 & \bf 8.42$^\dagger$ & \bf 9.97$^\dagger$  \\
\midrule
Single PW-HuBERT & \underline{48.7} & \underline{52.5} / \underline{67.3} & \underline{59.44} / \underline{63.51} & \bf 88.32 & 98.44 &  5.16 & 6.88  \\ 
Hierarchical PW-HuBERT & \bf 49.06 & \bf 55.3 / \bf 68.6 & \bf 61.28 / \bf 65.55 & \underline{88.25} & \bf 98.85 & \underline{6.55} &  \underline{9.02}  \\
\bottomrule
\end{tabular}
}
\label{tab:main}
\end{table*}

\vspace{-10pt}
\subsection{Hierarchical PW-HuBERT}
To further explore the synergy between phone-level units (i.e. HuBERT units) and word-level units (i.e. pseudo units from section~\ref{sec:target}) in SSL pre-training, we propose Hierarchical PW-HuBERT to unify the two training objectives illustrated in Fig.~\ref{fig:hierarchical}.
The architecture of Hierarchical PW-HuBERT is similar to Single PW-HuBERT. We add two extra transformer layers after HuBERT-Base model.
Compared with Single PW-HuBERT, we omit the weighted-sum layer and unfreeze the HuBERT model in Hierarchical PW-HuBERT. 

Prior work \cite{pasad2021layer} has demonstrated that higher-level information usually exists in deeper layers. 
Therefore, we inject objectives of different granularity on different layers.
Specifically, the model aims to predict frame-level targets on the 12\textsuperscript{th} layer and word-level targets on the 14\textsuperscript{th} layer, which aligns with the information scope of these layers. 

Similar to Single PW-HuBERT, the pseudo word-level targets $\mathbf{Z}^{\prime}$ are used to calculate $\mathcal{L}_{\mathrm{pw}}$.
Furthermore, we keep the original training loss in HuBERT~\cite{hsu2021hubert} as frame-level objective $\mathcal{L}_{\mathrm{frame}}$.
The overall training loss $\mathcal{L}$ is the combination of $\mathcal{L}_{\mathrm{pw}}$ and $\mathcal{L}_{\mathrm{frame}}$:
\begin{equation}
    \mathcal{L} = \mathcal{L}_{\mathrm{pw}} + \lambda \mathcal{L}_{\mathrm{frame}}
\end{equation}
where $\lambda$ is a hyper-parameter. 
It's worth mentioning that the masked positions in $\mathcal{L}_{\mathrm{pw}}$ and $\mathcal{L}_{\mathrm{frame}}$ are generated independently to increase the diversity of training input.

\vspace{-10pt}
\section{Experiment}
\vspace{-10pt}
\label{sec:experiment}

\subsection{Datasets}
To evaluate whether our model has a better understanding of higher level understanding of spoken languages, we run our models on a wide range of spoken language understanding benchmarks, including SLUE \cite{shon2022slue}, SLUE Phase-2 \cite{shon-etal-2023-slue}, SNIPS \cite{coucke2018snips}, and  ZeroSpeech 2021 \cite{nguyen2020zero}.  

\noindent\textbf{SLUE} is a benchmark for Spoken Language Understanding (SLU) evaluation on natural speech, which contains three tasks: sentiment analysis (SA), named entity recognition (NER), and automatic speech recognition (ASR). In this paper, we only focus on the SLU tasks: SA and NER.
SA aims to classify the sentiment of a speech utterance as having negative, neutral, or positive sentiment, and the metric employed is macro-averaged F1 score. 
NER involves predicting the name entities and their corresponding types in a given speech sequence. 
The metrics involve F1 and Label F1, which evaluate entities and corresponding types respectively.  Label F1 is useful to understand
model accuracy despite the possible misspelling and segmentation
errors in speech-to-text conversion.

\noindent\textbf{SLUE Phase-2} is a complement of SLUE, which introduces four additional SLU tasks. 
We only evaluate on the named entity localization (NEL) task as the codebase and datasets of other tasks have not been published yet. 
The goal of NEL is to locate the position of named entities in the audio by predicting their start and end times. 
The task evaluates performance via frame F1 and word F1 scores based on the overlap between the predicted and ground truth time segments. 
We follow the official training set and evaluate on the development set for both SLUE and SLUE Phase-2, since the test sets are not publicly accessible.

\noindent\textbf{ZeroSpeech 2021} is a benchmark designed to evaluate the model's performance on different linguistic levels, and we evaluate on the semantics task. 
Given pairs of spoken words, the task is to predict the similarity for each word pair and compare the correlation score between model prediction and human judgment.

\noindent\textbf{SNIPS} includes a collection of speech utterances along with corresponding annotations for slots and intents. We evaluate on the slot filling (SF) and intent classification (IC) tasks. SF aims to locate and classify entities within the input data. IC involves determining the user's intention given the speech utterance. For both tasks, we use F1 as the metric.

For SLUE, SLUE Phase-2, and SNIPS, we evaluate the tasks with SLUE toolkit\footnote{\href{https://github.com/asappresearch/slue-toolkit}{https://github.com/asappresearch/slue-toolkit}}. For ZeroSpeech 2021 semantics, we follow the instruction of ZeroSpeech2021 challenge\footnote{\href{https://zerospeech.com/challenge_archive/2021/04_instructions}{https://zerospeech.com/challenge\_archive/2021/04\_instructions}} to get the results.

\vspace{-5pt}
\subsection{Baselines}
\vspace{-5pt}
We compare our models with 3 different baselines: HuBERT \cite{hsu2021hubert}, VG-HuBERT \cite{peng2022word}, and a 14-layer variant of HuBERT - HuBERT$_{14}$. 
To investigate whether the improvement of our model comes from pseudo word-level targets rather than additional parameters, we add an extra baseline of a 14-layer HuBERT on all tasks.
Specifically, the training procedure is the same as the original HuBERT but with 14 transformer layers.

\vspace{-5pt}
\subsection{Implementation Details}
\vspace{-5pt}
For pseudo word-level target generation, we apply a threshold of 0.8 to the attention weights of layer 9 from VG-HuBERT$_3$\footnote{VG-HuBERT$_x$ refers to reinitializing the weights of the last $x$ transformer layers.} on LibriSpeech 100hr to get the word boundaries for our training dataset. 
Each target ID is obtained from running K-means clustering with 4096 clusters on the features extracted from the VG-HuBERT model and mean-pooled within the boundary. For frame-level target generation, we follow the same procedure of HuBERT~\cite{hsu2021hubert} to get the labels by applying the K-means algorithm on the 6th transformer layer from the previous iteration model.

As for the training configuration, the total training steps of Single PW-HuBERT and Hierarchical PW-HuBERT are 500k steps on a single GPU and 125k steps on 4 GPUs respectively. 
We apply a linear scheduler with warmup steps of 32k and a peak learning rate of 1e-4. 
The two additional transformer layers are randomly initialized and have the same dimensions as the previous layers.
\vspace{8pt}

\begin{table*}[ht]
\centering
\caption{The comparison with oracle setting, where we replace attention midpoint boundary with ground truth boundary and replace ID generated from clustering with ID from the BERT tokenizer.}
\vspace{5pt}
\resizebox{0.75\linewidth}{!}{
\begin{tabular}{c|cc|cc|c|cc}
\toprule
\multirow{2}{*}{\bf Architecture} & \multirow{2}{*}{\bf Boundary} & \multirow{2}{*}{\bf Target ID} & \multicolumn{2}{c|}{ \bf SLUE} & \bf SLUE 2 & \bf SNIPS \\
& & & \bf SA & \bf NER & \bf NEL & \bf SF \\
\midrule 
HuBERT & frame & cluster & 45.27 & 51.6 / 64.8 & 57.54 / 61.14 & \underline{88.2} \\
\midrule 
\multirow{3}{*}{Hierarchical} & ground truth & word ID & \underline{47.23} & 50.5 / 63.7 & 55.5 / 57.46 & 87.4 \\
 & ground truth  & cluster & 44.49 & \bf 56.6 / 70.8 & \bf 61.5 / \bf 66.40 & 88 \\
 & attention midpoint & cluster & \bf 49.06 & \underline{55.3} / \underline{68.6} & \underline{61.28} / \underline{65.55} & \bf 88.3 \\
\bottomrule
\end{tabular}
}
\label{tab:oracle}
\end{table*}

\vspace{-15pt}

\section{Results and Analysis}
\subsection{Main Results}

Table~\ref{tab:main} provides an overview of our experimental results. We can observe that both Single and Hierarchical PW-HuBERT show significant improvement over baseline models in SLUE and SLUE Phase-2, indicating the effectiveness of leveraging pseudo word-level targets. 
In SNIPS, the hierarchical PW-HuBERT consistently improves on both SF and IC task, although the improvements are relatively modest. 
In ZeroSpeech 2021 Semantics, the hierarchical PW-HuBERT yields better performance than HuBERT and HuBERT$_{14}$ on both librispeech and synthetic subsets, which demonstrates our model has a better ability to capture semantic information.
While our models exhibit lower performance than VG-HuBERT on both subsets, it is worth considering that VG-HuBERT has a similar training setup to ZeroSpeech semantics task, which calculates the similarities between embeddings. 
While this setup shows advantages on ZeroSpeech semantics task, it can be observed that it has a negative impact on general SLU tasks, including SLUE, SLUE Phase-2, and SNIPS.

Overall, our models' superior performance on SLUE, SLUE Phase-2, and SNIPS demonstrates their effectiveness in SLU downstream tasks. Furthermore, our improvements on ZeroSpeech 2021 semantics indicate the richness of word-level information within our pre-trained representations.


\vspace{-8pt}

\subsection{Comparison with Oracle Setting}

In this section, we discuss the effect of introducing ground truth word boundary and word token ID into our model architecture. 
To do so, we have two different settings, one is to directly replace the word boundary generated from VG-HuBERT with ground truth boundary and run K-means clustering as before, and the other is not only to substitute word boundary but also skip the clustering method by assigning word ID from the BERT tokenizer to each token. 
Notice that under these two settings, the model has to predict IDs in 4096 and around 30000 different labels respectively. 
For these two experiments, we employ the same hyper-parameters and model architecture as the Hierarchical PW-HuBERT model. 

As shown in Table~\ref{tab:oracle}, the improvement of using ground truth boundary is limited and unstable. We hypothesize that utilizing boundaries from attention weights provides the model with more precise locations to focus on. This is attributed to the attention-midpoint boundary typically narrower than the ground truth, consequently contracting more useful information for both K-means clustering and our model. Our findings contradict with \cite{Nguyen2022AreWB}, who show that LSTM models with boundaries predicted by DP-Parse or gold boundaries perform worse than LSTM models without any boundaries\footnote{We consider models training with phone unit from table 5 of \cite{Nguyen2022AreWB}, as it is similar to our setting.}. We think the disparity arises from the usage of word boundaries. Specifically, we aggregate the information within the word boundaries, whereas \cite{Nguyen2022AreWB} simply adds a $\langle \text{SPACE} \rangle$ token between words.


As for the worse performance on oracle boundaries and oracle word IDs, we suppose that IDs from clustering results have more semantic relations between each ID entry than ground truth IDs.

\subsection{Ablation Studies}

\subsubsection{The Effect of Frame-level Targets}
\label{sec:frame}
We investigate the effect of incorporating frame-level targets with pseudo word-level targets.
We conduct an analysis between Hierarchical PW-HuBERT and its counterpart without frame-level targets. 
The latter setting is equivalent to Single PW-HuBERT but the weighted-sum layer is omitted and the HuBERT weight is unfrozen.

The results in Table~\ref{tab:target} show that the performance consistently improves on almost all datasets when integrating frame-level targets. 
This suggests that leveraging the synergy between frame-level and word-level targets can further provide guidance for the training process.

\begin{table}[ht]
\Large
\caption{Ablation studies of hierarchical PW-HuBERT on frame-level supervision.}
\centering
\resizebox{\columnwidth}{!}{
    \begin{tabular}{cc|cccc}
        \toprule
        \multirow{2}{*}{ \bf  Architecture} & \bf HuBERT & \multicolumn{2}{c}{\bf SLUE} & \bf SLUE 2 & \bf SNIPS \\
        & \bf Targets & \bf SA & \bf NER & \bf NEL &  \bf SF \\
        \midrule
        \multirow{2}{*}{Hierarchical}  & \xmark & 44.94 &  53.6/67.9 & 59.4 & \bf 88.4 \\
         & \cmark & \bf 49.06 & \bf 55.3/68.6 & \bf 61.3 & 88.3 \\
         \bottomrule
    \end{tabular}
    }
    \label{tab:target}
\end{table}

\vspace{-15pt}
\begin{table}[ht]
\Large
\caption{Ablation studies of single PW-HuBERT on freezing HuBERT weights.}
\centering
\resizebox{\columnwidth}{!}{
    \begin{tabular}{cc|cccc}
        \toprule
        \multirow{2}{*}{ \bf  Architecture} & \multirow{2}{*}{\bf Freeze} & \multicolumn{2}{c}{\bf SLUE} & \bf SLUE 2 & \bf SNIPS \\
        & & \bf SA & \bf NER & \bf NEL &  \bf SF \\
        \midrule
        \multirow{2}{*}{Single}  & \xmark & 44.94 & \bf 53.6 / 67.9 & 59.44 / \bf 63.51 & \bf 88.4 \\
        & \cmark & \bf 48.7 & 52.5 / 67.3 & \textbf{59.48} / 62.56 & 88.3 \\
         \bottomrule
    \end{tabular}
    }
    \label{tab:freeze}
\end{table}

\vspace{-15pt}

\subsubsection{Freezing HuBERT Weights of Single PW-HuBERT}
In this section, we discuss the effectiveness of freezing HuBERT weights in Single PW-HuBERT. As shown in Table~\ref{tab:freeze}, freezing HuBERT weights and adding a weighted-sum layer yields comparable performance with the results obtained when HuBERT weights are frozen across all datasets. Notably, this approach has better training efficiency, which prompts us to adopt freezing HuBERT weights as our final method for Single PW-HuBERT.

\vspace{-10pt}
\section{conclusion}
\vspace{-5pt}
In this work, we address the challenge of insufficient semantic information in recent SSL models. 
We propose PW-HuBERT, a framework that incorporates pseudo word-level targets from a visually-grounded model into the training process without the necessity of speech-text paired data.
Our experiments on several SLU tasks show the effectiveness of our models in semantic understanding. 
Furthermore, we demonstrate that oracle boundary may not be optimal, as it potentially contains unuseful information. Finally, we show the benefits of joint training with frame-level targets and freezing HuBERT weights in the ablation study. 
In our future work, we plan to explore the impact of introducing intermediate-level units, such as syllable-level information, in speech SSL models and introduce the whole word masking strategy for PW-HuBERT.

\vspace{-8pt}

\section{ACKNOWLEDGEMENT}

This material is based upon work supported by the National Science Foundation under Grant No. 2238605. We thank the National Center for High-performance Computing (NCHC) in Taiwan for providing computational and storage resources.

\bibliographystyle{IEEEbib}
\bibliography{strings,refs}

\end{document}